\documentclass[aps,pra,twocolumn,showpacs,groupedaddress,floatfix]{revtex4}  
\usepackage{graphicx}  
\usepackage{dcolumn}   
\usepackage{bm}        
\usepackage{amssymb}   
\usepackage[export]{adjustbox}

\usepackage{sidecap}
\usepackage{subfigure}

\usepackage{amsmath}

\begin{document}



\title{Controlling electron localization of H$_2^+$ by intense plasmon-enhanced laser fields}

\author{I. Yavuz$^1$}
\author{M. F. Ciappina$^2$}
\author{A. Chac\'on$^3$}
\author{Z. Altun$^1$}
\author{M. Lewenstein$^{3,4}$}

\affiliation{$^1$Marmara University, Physics Dep. 34722, Ziverbey,
Istanbul, Turkey} \affiliation{$^2$Max-Planck-Institut f\"ur
Quantenoptik, Hans-Kopfermann-Strasse 1, D-85748 Garching,
Germany} \affiliation{$^3$ICFO-Institut de Ci\`encies
Fot\`oniques, The Barcelona Institute of Science and Technology,
08860 Castelldefels (Barcelona), Spain}
\affiliation{$^4$ICREA-Instituci\'{o} Catalana de Recerca i
Estudis Avan\c{c}ats, Lluis Companys 23, 08010 Barcelona, Spain}

\date{\today}

\begin{abstract}

We present a theoretical study of the wave packet dynamics of the
H$_2^+$ molecular ion in plasmon-enhanced laser fields. Such
fields may be produced, for instance,  when metallic nano-structures
are illuminated by a laser pulse of moderated intensity. Their
main property is that they vary in space on nanometer scales. We
demonstrate that the spatial inhomogeneous character of these
plasmonic fields leads to an enhancement of electron localization,
an instrumental phenomenon that controls molecular fragmentation.
We suggest that the charge-imbalance induced by the
surface-plasmon resonance near the metallic nano-structures is the
origin of the increase in the electron localization.

\end{abstract}

\pacs{42.65.Ky, 78.67.Bf, 32.80Ee}
\maketitle

\section{Introduction}

Studies of atomic and molecular quantum dynamics are in the center
of interests of contemporary atomic, molecular and optical
physics. There are various ways to induce such dynamics, but a
very distinct one is  to expose the atomic/molecular systems to
an intense and coherent electromagnetic radiation. As a
consequence of this coupling, new and diverse phenomena occur.
Amongst the plethora of processes which take place, the two most
prominent ones are the high-order harmonic generation (HHG) and
the above-threshold ionization (ATI). They both lie at the core of
the so-called attosecond
physics~\cite{corkum2007attosecond,FerencReview}. The
quasi-classical picture of these two phenomena relies on the so-called three-step or simple man's
model~\cite{corkum1993plasma,lewenstein1994theory}. Briefly, this
approach can be summarized listing the subsequent steps: (i)
tunnel ionization due to the intense and low frequency laser
field; (ii) acceleration of the free electron by the laser
electric field, and (iii) re-collision with the parent ion after
the temporally oscillating laser electric field reverses the
direction of the electronic motion. In HHG the electron recombines
with the remaining ion-core and the excess of energy is converted
into a high energy photon~\cite{FerencReview}. On the other hand,
if the electron is re-scattered by the atomic potential, it gains
even more kinetic energy and contributes to the high energy parts
of the ATI spectrum~\cite{milosevic2006review}.

Commonly, the laser ionization process is understood by invoking
the quasi-static tunnel ionization picture. In this approach in
every time (short) interval the laser electric field is considered
as a static electric field. This assumption is valid for low
photon frequencies -- long wavelengths, and the atomic or
molecular electron is considered to quickly tunnel out through the
barrier created by the combined potentials of the laser electric
field and the attractive Coulomb atomic/molecular potential. The
main consequence of this description is that the ionization rate
presents a maximum whenever the barrier becomes the thinnest,
which correlates with the electric field maxima. This prediction
appears to be also valid in the so-called non-adiabatic tunnel
ionization, i.e.~when the laser electric field is considered to
change significantly, while the electron is escaping from the
attractive potential of the atomic or molecular core. For these
cases the ionization rate exhibits a single maximum during each
half-cycle of the laser electric field
oscillation~\cite{yudin2001adk}.

When molecules are used as driven media, the above cited
assumptions should be revised, considering that there can be
multiple bursts of ionization within a half-cycle of the laser electric
field~\cite{takemoto2010,takemoto2011}. By using different
numerical models and simulations, it was confirmed that these
bursts are related to the effect of transient electron
localization (EL) at one of the heavy nuclei of the molecule on a
sub-fs time scale~\cite{kawata1999,he2008,takemoto2011PRA}.
Generally, a sub-cycle oscillation of the electron density occurs
after the molecular ion has been stretched to intermediate
internuclear distances, and it is due to a trapping of the
electron population in a pair of so-called charge-resonance (CR)
states~\cite{mulliken1939,paulingpnas}. For the case of a simple
H$_2{^+}$ molecular ion, they are the energetically lowest
$\sigma_g$ and $\sigma_u$ states. It is likely, however,  that
both the multiple ionization bursts, and the CR appear in other
more complex molecules as well. In addition, 
EL appears to be the responsible of the strongly enhanced ionization rate, observed 
for stretched molecules beyond its equilibrium
internuclear separation~\cite{TamarPRL}, and, moreover, it was shown EL can be manipulated, joint with the control of photoabsorption/photodissociation, by using different alignment techniques (see e.g.~\cite{TamarRMP}).

In the theoretical modeling of conventional strong laser-matter
interaction, the main assumption is that both the laser electric
field ($E(\mathbf{r},t)$) and its vector potential associated
($A(\mathbf{r},t)$) are spatially homogeneous in the region where
the electron develops its motion and only their time dependence is
considered, i.e.~$E(\mathbf{r},t)=E(t)$ and
$A(\mathbf{r},t)=A(t)$. This is a legitimate assumption since the
fields  change at most on the scale of the wavelength ($800-3000$
nm), while the typical size of a laser focus is between several
tens to a couple of hundreds microns ($10^{-6}$ m). These scales
have to be compared with the size of the electronic ground states
(say of the order of Angstroms ($10^{-10}$ m)), and the size of the
typical electron excursions, estimated classically using
$\alpha=E_0/\omega_0^2$; even these sizes remain sub-wavelength
and do not reach more than tens of nm ($10^{-9}$ m) for longer wavelengths and
higher laser intensities (note that $\alpha\propto \lambda_0^2$,
where $\lambda_0$ is the wavelength of the driven laser and
$E_0=\sqrt{I}$ where $I$ is the laser
intensity)~\cite{FerencReview}. On the contrary, the fields
generated using surface plasmons are spatially dependent on a
nanometer scale. By exploiting the surface plasmon resonance
(SPR), locally enhanced electric field can be induced around gold
bow-tie nano-antennas. The enhanced field boosts up the low
incoming laser intensity, specified in the 10$^{11}$ W cm$^{-2}$
range, by more than 30 dB, which becomes then strong enough to
exceed the threshold laser intensity for HHG generation in noble
gases~\cite{kim2008high}. The pulse repetition rate, typically in
the MHz domain, remains unaltered without any extra pumping or
cavity attachment; this is one of the main advantages of this
setup. From a theoretical viewpoint, plasmonic fields open a wide
range of possibilities to enhance and/or shape spectral and
spatial properties of the incoming
fields~\cite{park2011plasmonic,choi2012generation,pfullmann2013bow,park2013generation}.
A peculiar property of this plasmonic fields is that the enhanced
laser electric field is not spatially homogeneous, on the scales
and in regions of comparable dimensions with the size of
the electronic excursion $\alpha$, where the electron dynamics takes
place.  Consequently, significant changes in the laser-matter
processes arise.

There has been a remarkable theoretical activity on this subject
recently~\cite{Husakou11A,Husakou11OE,Marcelo12OE,Marcelo12A,Marcelo12AA,Marcelo12AAA,Marcelo12JMO,yavuz2012generation,Jose13,
yavuz2013gas,Marcelo13A,Marcelo13AR,Marcelo13AP,Marcelo13AA,Marcelo13LPL,Marcelo14,Marcelo14EPJD,
Marcelo15CPC,Marcelo15,Husakou14A,Ebadi14A,Fetic12,Luo13,Feng13,Wang13,Lu13A,He13,Zhang13,Luo13JOSAB,
Cao14,Cao14a,Feng15,Yu15}. In most of the contributions, however,
only the HHG and ATI processes in atoms were studied and analyzed
exclusively. Only recently investigations of HHG in simple
molecules, H$_2^+$, driven by plasmonic fields were
presented~\cite{yavuz2015h2+,FengPRA}. In the present paper we focus on the
question to what extent plasmonic fields could configure a novel
and reliable tool to control molecular dynamics, and in particular
the electron localization (EL). As mentioned above, these fields
are spatially inhomogeneous and thus they offer to our disposal a
new degree of control, which could help us to obtain an even more
precise manipulation of the electron and molecular dynamics at a
sub-cycle time scale.

Our article is organized as follows. In Sec.~II, we describe the
methods, including here a brief description about the set up that
could be utilized for the generation of plasmonic enhanced fields
and its main characteristics. Once we have defined our theoretical
model and observables of interest, we then analyze them and
discuss their properties and implications in Sec.~III. Finally, in
Sec.~IV, we end the paper with conclusions and a brief outlook.

\section{Methodology}

\subsection{Field-enhancement by the nano-structure}

A typical set up including both the metal bow-tie nano-antenna and the driven H$_2^+$ molecule is shown in Fig.~1 (for more details about the fabrication see e.g.~\cite{kim2008high}). The metal nano-structure takes the form of two triangular-shaped pads made of gold with a gap between their apexes (top panel). A planar substrate with a dielectric constant $\epsilon_s=2.0$ supports them. In the diagram, $h$ is the height, $t$ is the thickness of the metal and $g$ defines the gap between the tips. In the present simulations, we use $h=100$ nm, $t=40$ nm and $g=20$ nm. We note that the chosen geometry parameters of the nano-antennae do not correspond to an optimum field-enhancement, i.e.~the nano-antenna is not fully resonant with the laser wavelength, set at $\lambda=800$ for the finite-difference time-domain (FDTD) simulations (see e.g.~\cite{Marcelo12OE}), but these values are found to be sufficient to understand the underlying physics of the wave packet dynamics of the H$_2^+$ molecular ion near the metallic nano-structure. In order to mimic more realistic situations, the curvature radii of the tips are set to be $4$ nm. The spatial profile of the field-enhancement around the bow-tie is determined using the MEEP code~\cite{oskooi2010meep}, that is based on the FDTD approach. The field-enhancement in the gap along the $z$-axis is shown in the bottom panel of Fig.~1, where the spatial profile is normalized to the intensity enhancement at the center ($z=0$). We observe that the laser electric field peak amplitude is enhanced roughly by a factor of 2.5 near the metal tips compared with the center value, corresponding to a 4 dB of increase in the laser intensity.
  
 \begin{figure}[!h]
\vspace{-32.0mm}
\centering
\begin{minipage}{\linewidth}
  \includegraphics[width=0.2\linewidth,center,angle=-90]{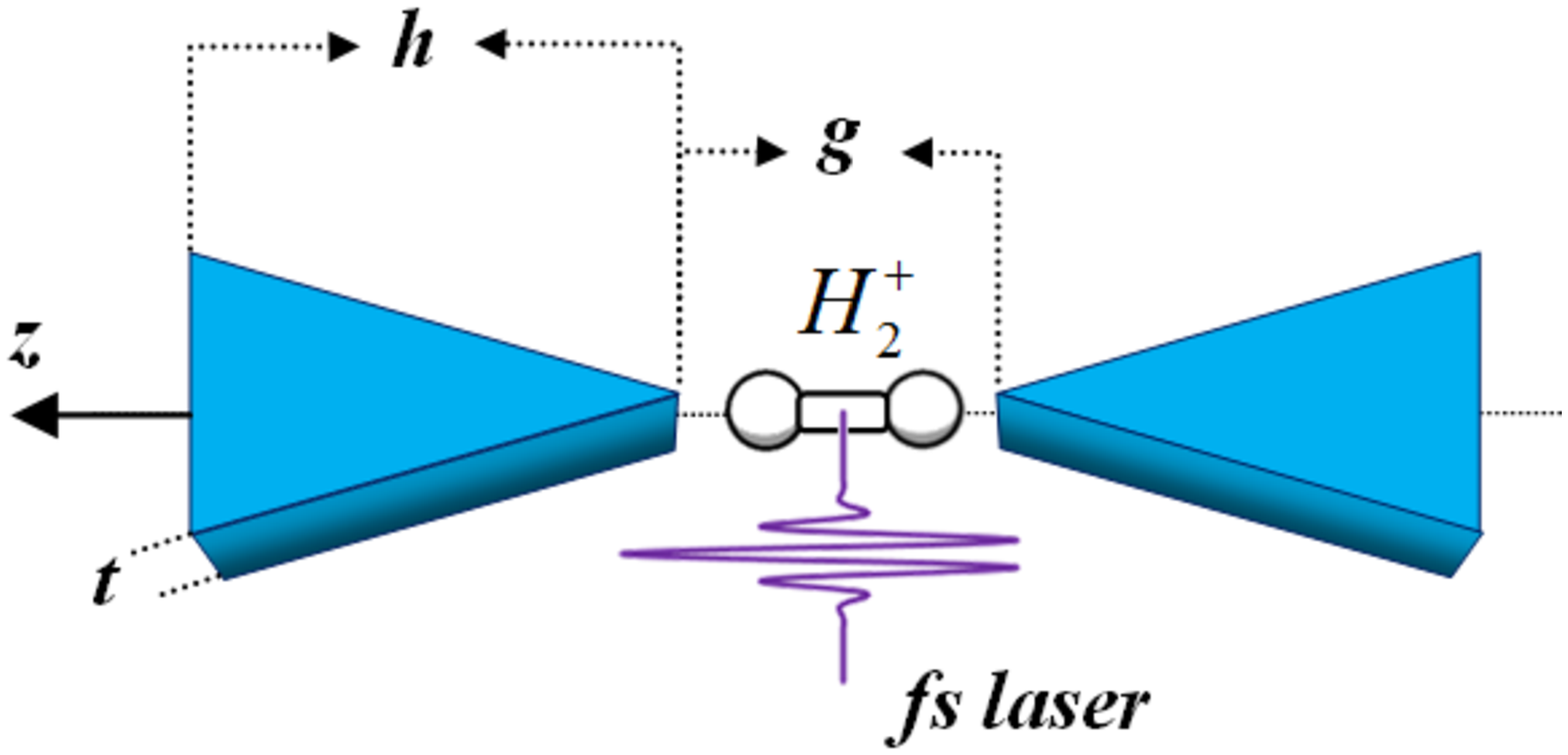}
    \label{img1}
\end{minipage}
\hspace{.0\linewidth}
\vspace{-10.0mm}

\begin{minipage}{\linewidth}
  \includegraphics[width=0.7\linewidth,center]{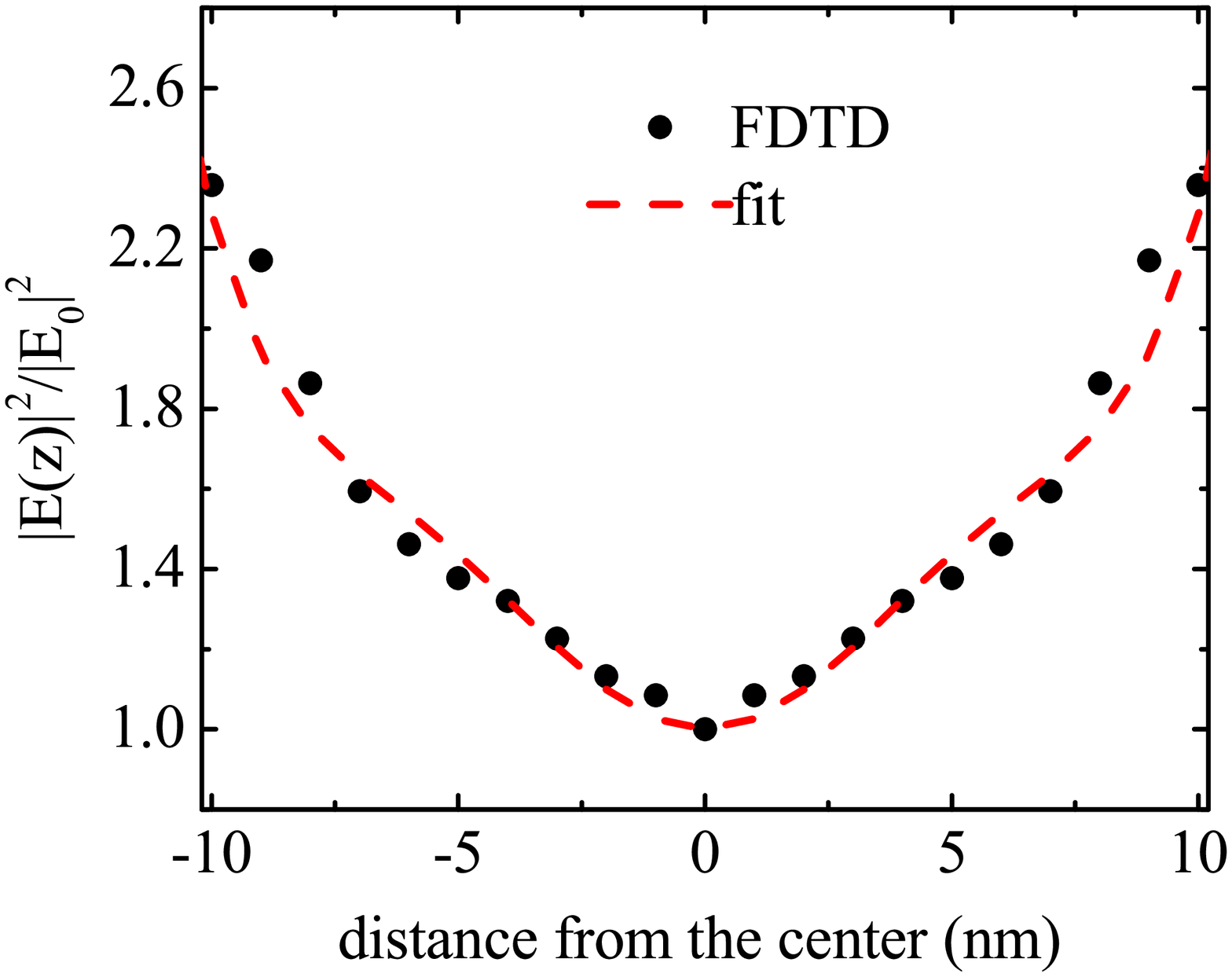}
  \caption{(color online) Top: Typical geometry parameters of the bow-tie shaped gold nano-antennae
  considered in our study. Here, we take $h=100$ nm, $t=40$ nm and $g=20$ nm. The curvature radii of the tips
  are taken as $4$ nm. The bond distance $R$ of the H$_2^+$ molecule is visually exaggerated for clarity. Bottom: The spatial
  profile of the field-enhancement in the gap, along $z$-axis, obtained from FDTD simulations. Circles are the
  actual enhancement determined by FDTD and solid line is a polynomial fitting, as described in the text.
  Note that, the laser electric field peak amplitude is enhanced roughly by a factor of 2.5 near the metals compared with the center value, corresponding to a 4 dB of
  increase in the laser intensity.}
  \label{img2}
\end{minipage}
\end{figure}

 \vspace{-7.0mm}
 \subsection{Numerical solution of time-dependent Schr\"odinger equation (TDSE) for the H$_2^+$ molecule}

 The numerical solution of the TDSE for the interaction of a linearly polarized laser
 field with a H$_2^+$ molecule, in reduced dimensions, is considered as,

\begin{equation} \label{eq1}
\begin{split}
 i\frac{d}{dt} & \Psi (z,R,t)   =  [ -\frac{1}{2}\frac{\partial^2 }{\partial z^2} -\frac{1}{2\mu_p} \frac{\partial^2 }{\partial R^2} \\
& +  V_e(z,R) + V_n(R)+V_L(z,t)  ] \Psi (z,R,t) .
\end{split}
\end{equation} where $\mu_p=m_p/2$ is the reduced-mass of the two nuclei. Here,
the potentials $V_e(z,R)$ and $V_n(R)$ are the electron-nuclei attraction and nucleus-nucleus repulsion terms,
respectively. The explicit forms of these potentials in reduced dimensions are given below~\cite{kulander1996model}:
\begin{equation} \label{eq2}
\begin{split}
V_e(z,R) &=-\frac{a}{\sqrt{(z+R/2)^2+b(R)}}-\frac{a}{\sqrt{(z-R/2)^2+b(R)}}, \\
\end{split}
\end{equation}
and
\begin{equation} \label{eq3}
\begin{split}
V_n(R)   &=\frac{1}{R}.
\end{split}
\end{equation}

In Eq.~(\ref{eq2}) $a=0.251$ is a scaling parameter and $b(R)$ is
a function introduced to exactly reproduce the 3D potential energy
curve of the $1\sigma_g$ state of the H$_2^{+}$ molecule. The
laser-molecule  interaction term $V_L(z,t)$ in the dipole
approximation can be written as:

\begin{equation} \label{eq4}
\begin{split}
V_L(z,t) &= -z E(z,t)   \\
         &= -z E(t)[1+s \kappa(z)].
\end{split}
\end{equation}
where it is assumed that the laser electric field $E(z,t)$ is now
function explicitly of both time and space. In Eq.~(\ref{eq4})
$\kappa(z)=\Sigma_i c_i z^i$ is a polynomial series that
represents the functional form of the plasmonic field, determined
by fitting it to the data obtained from FDTD simulations (see
Fig.~1, bottom panel) and $s$ is a switch function taking values
$s=0,1$. $s$ is used to turn the field inhomogeneity on or off, as
discussed later. Note that the inter-nuclear axis of the H$_2^+$
molecule is placed as to coincide with the axis passing through
the edges of the nano-structure element (see Fig.~1, top panel).

The field-free form of Eq.~(\ref{eq1}),

\begin{equation} \label{eq5}
\begin{split}
\left [ -\frac{1}{2}\frac{\partial^2 }{\partial z^2}  -\frac{1}{2\mu_p}  \frac{\partial^2 }{\partial R^2} + V_e(z,R) +  V_n(R)  \right ] &  \Psi (z,R)  \\
= E  & \Psi (z,R) .
\end{split}
\end{equation}
is numerically solved based on the Born-Oppenheimer (BO)
approximation. When we apply the BO approximation to
Eq.~(\ref{eq5}), its solution is expressed as follows:

\begin{equation} \label{eq6}
\Psi (z,R,t=0) = \phi_e(z,R) \psi_n (R)
\end{equation}
where $\phi_e(z,R)$ is a set of electronic wavefunctions for fixed
values of $R$ and $\psi_n (R)$ is the nuclear wave function. Both
$\phi_e(z,R)$ and $\psi_n(R)$ are calculated from the eigenvalue
equations:

\begin{equation} \label{eq7}
\left [ -\frac{1}{2}\frac{\partial^2 }{\partial z^2} + V_e(z,R) \right ]   \phi (z,R,t)  = E_e(R) \phi (z,R,t),
\end{equation}

and

\begin{equation} \label{eq8}
\left [ -\frac{1}{2\mu_p}  \frac{\partial^2 }{\partial R^2} + E_e(R) + V_n(R)\right ]  \psi (R) = E \psi (R),  \\
\end{equation}
respectively.

The field-free solutions of Eq.~(\ref{eq7}) and (\ref{eq8}) for
H$_2^+$ give an equilibrium bond distance $R=2.0$ a.u. The
ionization potential $I_p$ at the equilibrium $1\sigma_g$ state is
found to be $|E_e(R=2.0)|=$ $30.0$ eV, i.e.~the actual value $I_p$
of the H$_2^+$ molecule is indeed reproduced.

The time-dependent part of the electric field in Eq.~(\ref{eq4})
is taken as $E(t)= E_0 f(t) \cos(\omega_0 t)$. $E_0$ and
$\omega_0$ are the peak amplitude [$E_0$ (a.u.) $= \sqrt{I/I_0}$ and
$I_0 = 35.1$ PW cm$^{-2}$] and the frequency of the driving laser
electric field, respectively. $f(t)$ is a flat-top, 10-cycles long
pulse envelope with half-cycle ramp up/down (total time duration
27 fs). During the simulations both the electronic and nuclear
wave functions are multiplied by mask functions of the form
$cos^{1/8}$ in each time step in order to avoid spurious
reflections at the boundaries~\cite{krause1992calculation}.

By using the time-dependent wave-function $\Psi(z,R,t)$ of
Eq.~(\ref{eq1}) it is then possible to compute a set of physical
quantities of interest, namely:

\noindent
(i) the time-dependent norm $N(t)$
\begin{equation} \label{eq9}
N(t)=\int_{0}^{\infty } dR\int_{-\infty}^{\infty } dz |\Psi(z,R,t)|^2,
\end{equation}
(ii) the ionization probability $P_{ion}(t)$,
\begin{equation} \label{eq10}
P_{ion}(t)=1-N(t),
\end{equation}
(iii) the dissociation channels through upper $P_{+}(t)$ or lower $P_{-}(t)$ nuclei
\begin{equation} \label{eq11}
P_{\pm}(t)=\int_{R_c}^{\infty } dR\int_{0}^{\pm z_c } dz |\Psi(z,R,t)|^2,
\end{equation}
(iv) the dissociation probability $P_{dissoc}(t)$
\begin{equation} \label{eq12}
P_{dissoc}(t)=P_+(t) + P_-(t),
\end{equation}
and (v) the asymmetry parameter $A(t)$
\begin{equation} \label{eq13}
A(t)=P_+(t) - P_-(t).
\end{equation}
The integration limits in Eq.~(\ref{eq11}) $R_c$ and $z_c$ are
taken as 10 a.u. The asymmetry parameter $A(t)$ determines the
degree of localization in either of the heavy nuclei upon
dissociation. Here, $A(t)>0$ or $A(t)<0$ refer to a high degree of
localization on the upper or the lower nuclei, respectively. For
$A(t)=0$, the dissociative wave-packet is evenly distributed over
both nuclei or no dissociation occur at all.

Finally, the time-dependent expectation value of the internuclear distance $R(t)$ is calculated by the following expression:
\begin{equation} \label{eq14}
\langle R(t) \rangle=\frac{1}{N(t)} \int_{0}^{\infty } R dR\int_{-\infty}^{\infty } dz |\Psi(z,R,t)|^2.
\end{equation}
This quantity allows us to monitor the time dynamics of the molecular dissociation.

\section{Results}

 \begin{figure}[!ht]
\centering

\vspace{-20.0mm}

\begin{minipage}{\linewidth}
  \includegraphics[width=0.7\linewidth,center,angle=270]{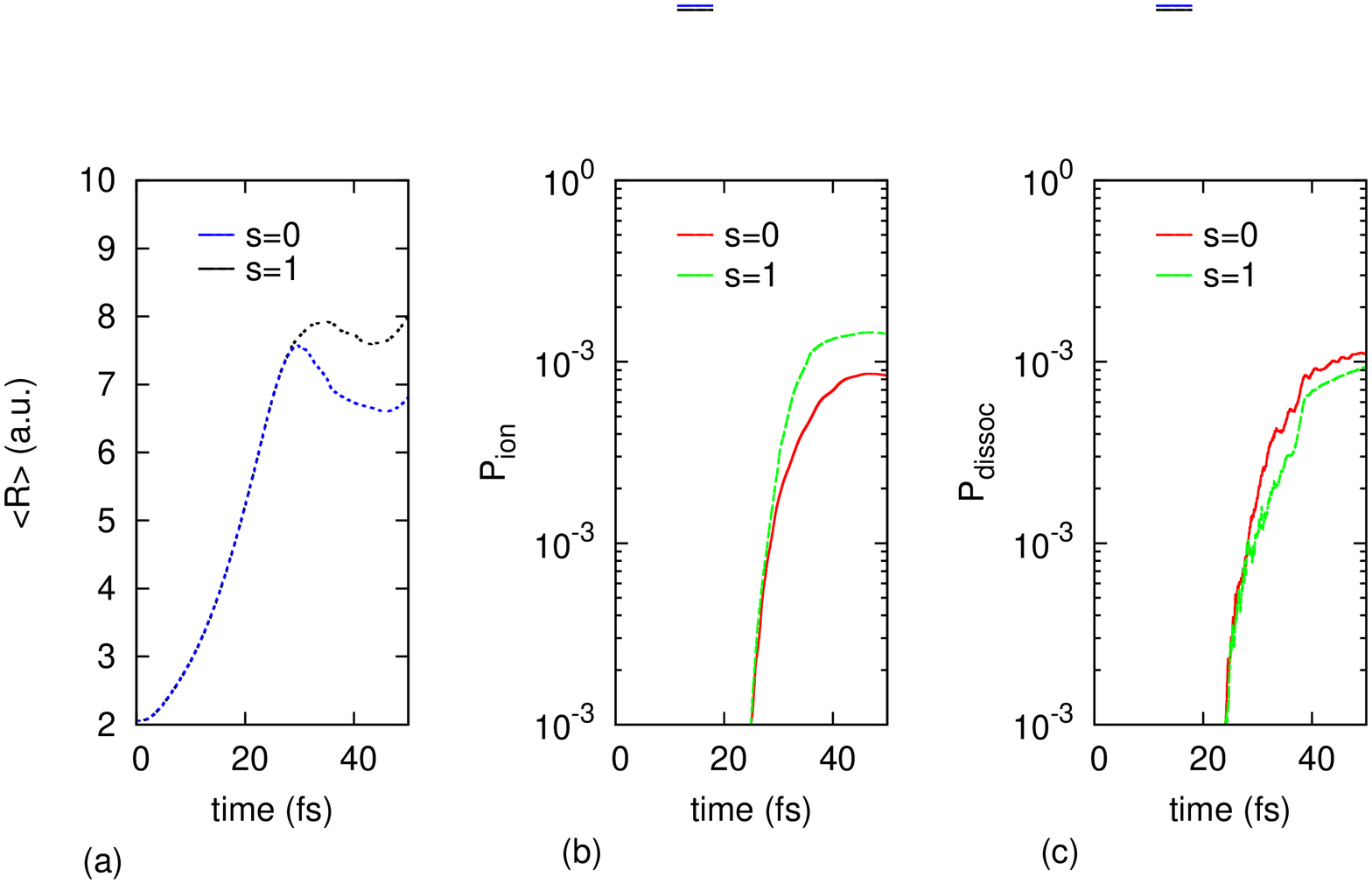}
  \vspace{-40.0mm}
    \label{img3}
\end{minipage}
\hspace{.0\linewidth}
\vspace{-20.0mm}
\begin{minipage}{\linewidth}
  \includegraphics[width=.35\linewidth,center,angle=-90]{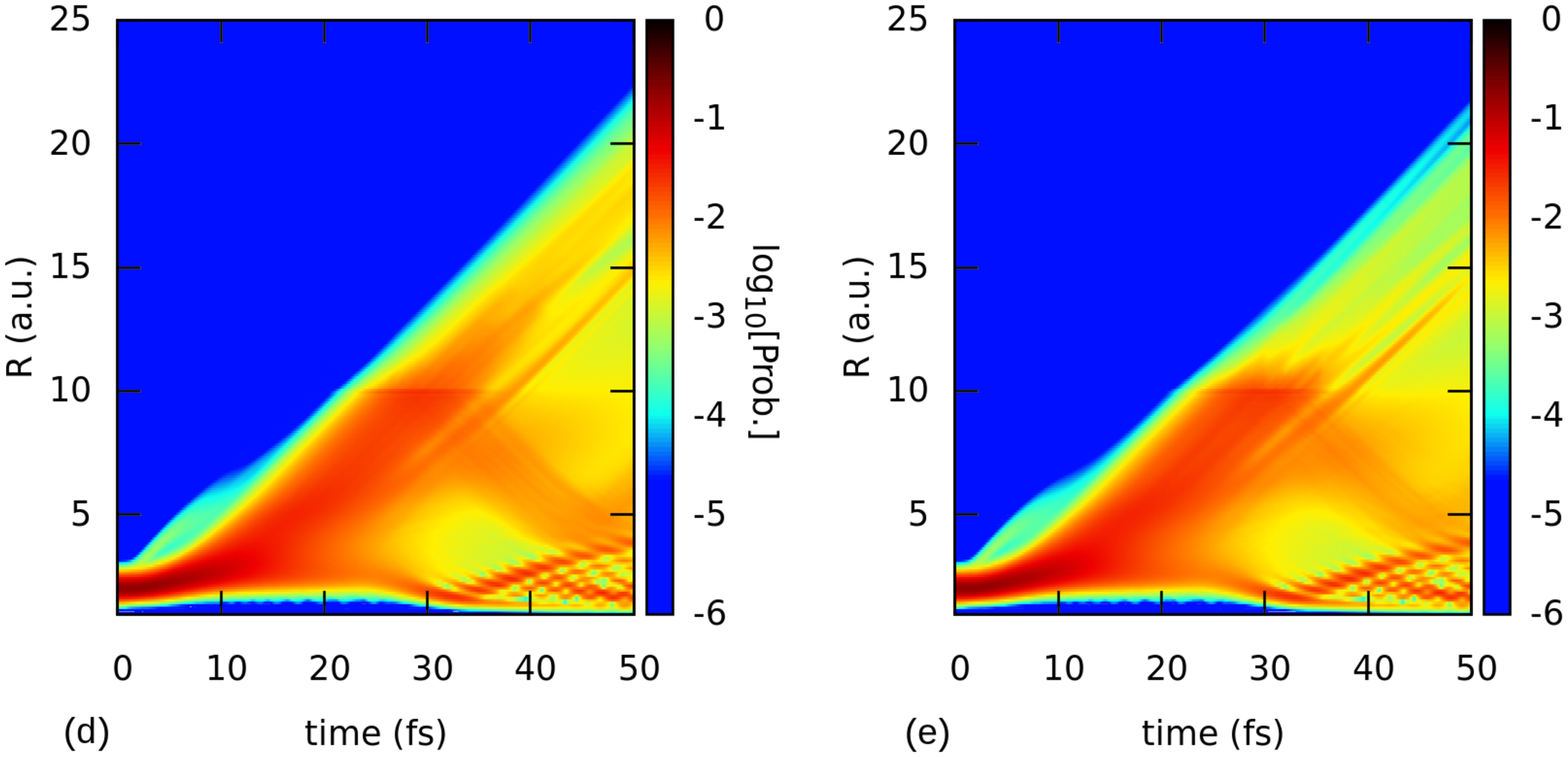}
    \label{img4}
\end{minipage}
\vspace{-20.0mm}
\hspace{.0\linewidth}
\begin{minipage}{\linewidth}
  \includegraphics[width=.35\linewidth,center,angle=-90]{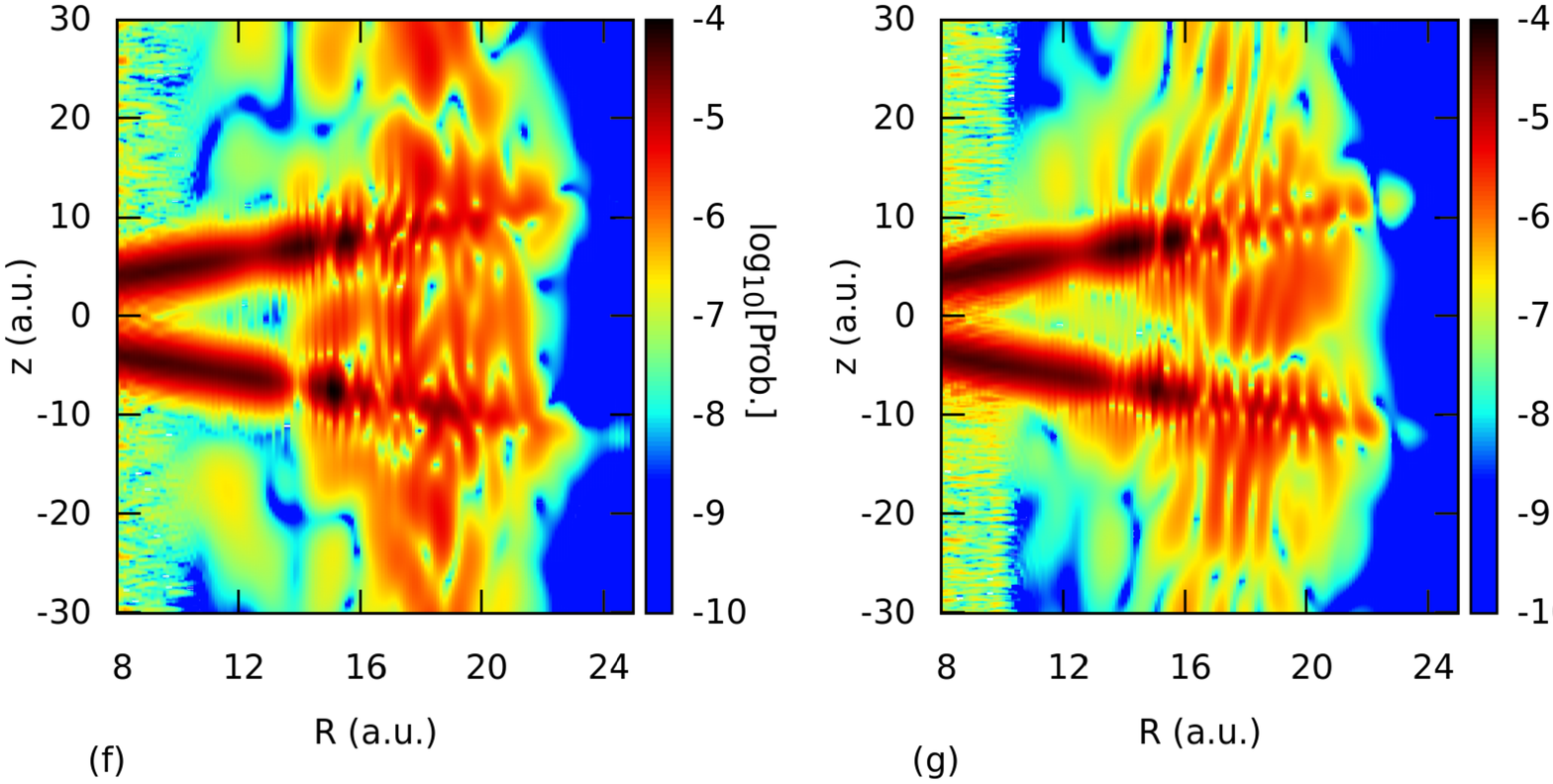}
  \vspace{-20.0mm}
  \caption{(color online) (a) Temporal variation of $R$, i.e.~$\langle R(t)\rangle$, (b) ionization probability and (c)
  dissociation probability of the H$^+_2$ for $s=0$ and $s=1$ (see Eq.~(\ref{eq4}) for more details). (Center)
  Time dependence of $R$-resolved probability distribution $|\Psi (R,t)|^2$ of H$^+_2$ for $s=0$ (left) and $s=1$ (right). (f) and (g)
  show the snapshots of the wave-packet distributions of $s=0$ and $s=1$, when the wave-packet is relaxed for a
  sufficient amount of time after the pulse ends. The laser intensity is $I=300$ TW cm$^{-2}$ and $\lambda = 800$ nm.
  The pulse comprises 10 total cycles (27 fs) and we use a flat-top pulse envelope with half-cycle ramp up/down.
  }
\end{minipage}
\end{figure}

In this section, we explore the influence of the spatial
inhomogeneous character of the plasmonic field on the
dissociation/ionization dynamics of H$_2^+$ molecular ion by
comparing results for both conventional and spatial inhomogeneous
fields. As stated otherwise, we consider that the centroid of
H$_2^+$ coincides with the center of the gap of the metallic
nano-structure, as shown in Fig.~1, and the molecule is initially
in the ground electronic $1\sigma_g$ and vibrational $v=0$ states.

Here, one can think the field spatial inhomogeneity as a potential landscape whose strength
is enhanced as we move away from the gap center to the metallic surfaces (see Fig.~1). In contrast, an
homogeneous field is independent of space, thus, constant in the region where the electron dynamics take place. In other words,
the plasmonic character of the laser electric field is effective when the wave-packet is released from the bound state and
spread across distant regions. Accordingly, there is a strong correlation between the ionization rate and the degree
of field spatial inhomogeneity. On the other hand, since opposing charges are confined in opposing sides of the nano-structure
due to SPR~\cite{kim2008high}, freed electrons of target atoms/molecules would experience diverse (repulsive or attractive) forces
depending on the direction of wave-packet's propagation.

Figures 2(a)-2(g) show time-dependent wave packet properties of
the H$_2^+$ molecular ion in a laser field with and without a spatial
inhomogeneous character ($s=0$ and $s=1$, respectively). The laser
intensity and the wavelength of the laser field are fixed at
$I=300$ TW/cm$^2$ ($3\times 10^{14}$ W/cm$^2$) and $\lambda=800$ nm, respectively. Firstly, in
Fig.~2(a) we show the time-dependent variation of $R$, i.e.~$\langle R(t)\rangle$. Since,
the laser field is switched off at $t=27$ fs, it is clear that
the plasmonic-laser field is effective when the wave-packet begins to relax
resulting in a slightly higher bond elongation. It is also evident
from Fig.~2(b) that the field inhomogeneity emerges when
ionization reaches a certain level (again almost immediately after
the pulse is over), which is $P_{ion} \sim 10^{-2}$. In addition,
the ionization probability increases roughly by a factor of two
after reaching a, sort of, limiting value. However, as shown in
Fig.~2(c), the dissociation probability is only slightly lower
for $s=1$ (inhomogeneous) than for $s=0$ (homogeneous), suggesting
a more direct electron ionization channel.

Figures 2(d)-2(e) show the time-variation of the nuclear probability
density for $s=0$ and $s=1$, respectively. In the bound region
(i.e.~for $R<10$ a.u.), $s=0$ and $s=1$ have similar profiles,
however the dissociation region (i.e.~when $R>10$ a.u.) is
slightly more occupied in the case of $s=0$, consistent with
Fig.~2(c).

Finally, Figs.~2(f)-2(g) show the electron-nuclear coordinate maps
when the wave-packets are relaxed for a sufficient amount of time
after the pulse is turned-off. Comparing with the homogeneous
case, and in the dissociation region, in particular for $R>14$
a.u., the wave-packet is much more localized on both nuclei for
the case of the plasmonic field. This is a significant outcome
showing the control the plasmonic character of the field has on
EL. In order to quantify this last
asseveration, using Eq.~(13), we find that, after the wave-packet
is relaxed, the asymmetry parameter is $A=-7.2\times 10^{-4}$ for
$s=0$ and $A=1.3\times 10^{-2}$ for $s=1$. Thus the asymmetry
(localization) for the case of the plasmonic field is roughly 20
times larger than the conventional case.

 \begin{figure}[!ht]
  \includegraphics[width=1.0\linewidth,center]{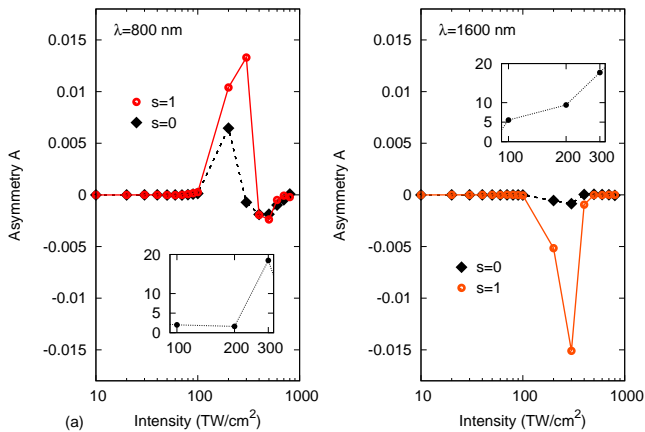}
  \caption{(color online) Variation of the asymmetry parameter ($A=P_+ - P_-$) parameter of H$^+_2$ as a function of plasmonic
  field-intensity for $s=0$ and $s=1$ (see Eq.~(4) for details). $A$ values are calculated after pulse is turned-off and then the
  system is left to relax for a sufficient amount of time. We have performed calculations for two different wavelengths;
  (a) $\lambda=800$ nm and (b) $\lambda=1600$ nm. Insets show the absolute asymmetry parameter enhancement, i.e.~the absolute ratio between $A$ of $s=1$ and $s=0$, in the region $I=100-300$ TW/cm$^2$. The temporal and spatial profile of the laser electric field is the same as in Fig.~1.}
  \label{img5}
\end{figure}

The intensity dependence of the $relaxed$ asymmetry parameter $A$
of wave packet dissociation at $\lambda=800$ and $1600$ nm is
shown in Fig.~3. For $s=0$, $A$ is enhanced in the intensity
region $100-300$ TW/cm$^2$ for both $\lambda=800$ and $1600$ nm.
For weak fields, $A$ is nearly zero due to the low wave packet
dissociation rates. For strong intensities, however, direct wave
packet ionization may occur, which also reduces dissociation.
Thus, there is an intermediate intensity region ($100-300$
TW/cm$^2$ in our case) such that dissociation reaches a maximum,
so is the asymmetry parameter $A$. On the other hand, comparing
with that of $\lambda=800$ nm, $A$ is lower in the intermediate
region ($100-300$ TW/cm$^2$) for $\lambda=1600$ nm. This is
attributed to faster vibrational motions induced by longer
wavelengths, causing an increase in the wave packet ionization.
When the molecule is placed in a plasmon-enhanced laser field
($s=1$), the asymmetry parameter $A$ is nearly zero in the weak
and strong intensity regions, similar to the case of $s=0$. On the other hand, in the
intermediate intensity region, $A$ is dramatically
enhanced for both $\lambda=800$ and $1600$ nm. In numbers, for
plasmonic fields at $\sim I=200-300$ TW/cm$^2$ the asymmetry
parameter $A$ is enhanced by a factor of $5\sim 20$ compared with the
conventional case for both $\lambda=800$ nm (see Fig.~3(a)) and
$\lambda=1600$ nm (see Fig.~3(b)).

\begin{figure}[!ht]
\vspace{-35mm}
\centering
\begin{minipage}{\linewidth}
  \includegraphics[width=0.35\linewidth,center,angle=-90]{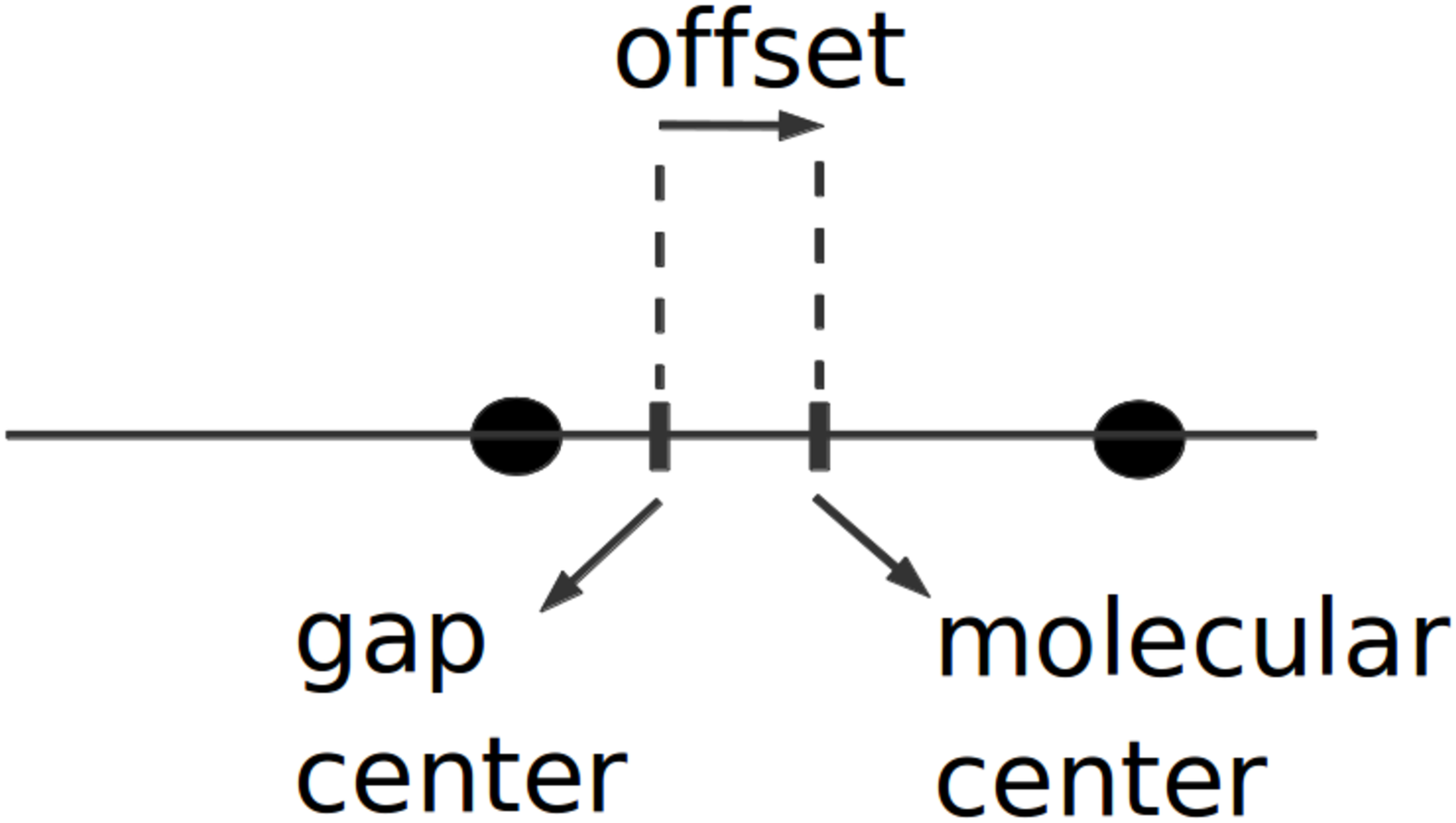}
   \label{img6}
\end{minipage}
\vspace{-35mm}
\hspace{.0\linewidth}
\begin{minipage}{\linewidth}
  \includegraphics[width=1.1\linewidth,center]{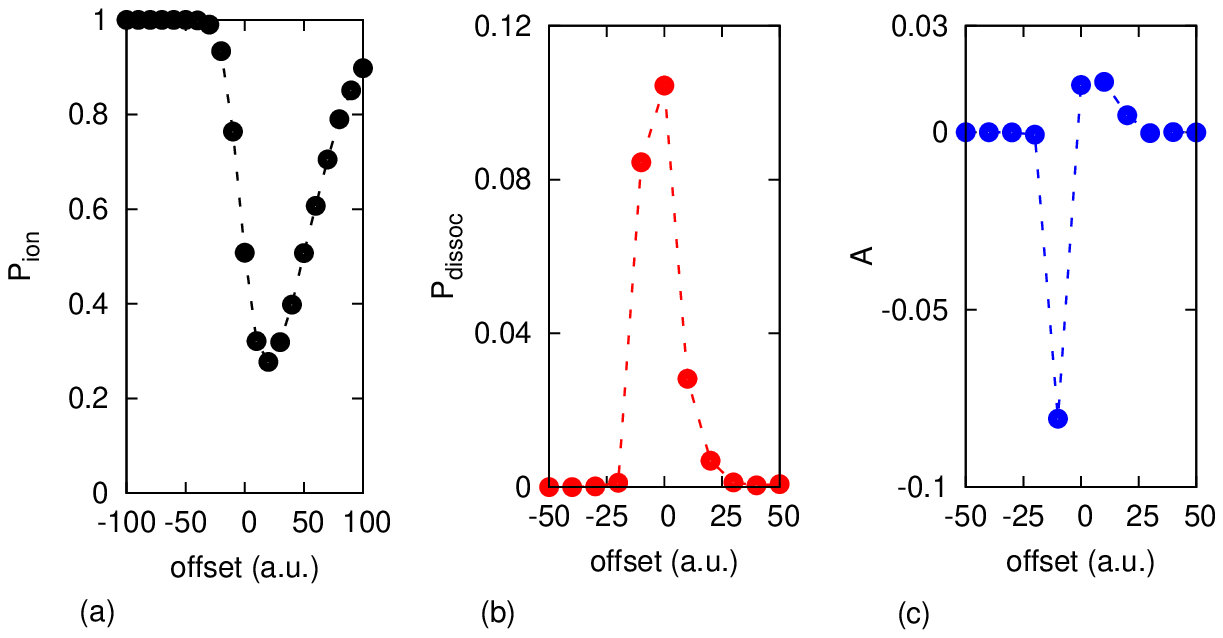}
\vspace{-10.0mm}
  \caption{(color online) Top: An illustration showing the gap center of the nano-structure element,
  molecular center (centroid) of H$_2^+$ and offset with respect to the gap center. Full circles illustrates the
  positions of the nuclei. (a) ionization probability, (b) dissociation probability and (c) asymmetry parameter,
  $A=P_+ - P_-$, as a function of the centroid offset. $P_{ion}$, $P_{dissoc}$ and $A$ are calculated after pulse
  is turned-off and then the system is left to relax for a sufficient amount of time. The laser field parameters are the same as in Fig.~2.}
\end{minipage}
\end{figure}

We argue that a charge-imbalance, in the nano-gap region, induced
by the plasmonic field is the responsible for the dramatic
enhancement in the asymmetry parameter $A$. So far, we assume that
the centroid of the molecule and gap center coincides, but in a
real experiment we could have molecules randomly distributed in
the nano-gap region. Thus one could ask, what happens if we
displace the molecule from the gap center to the left or right regions
along the $z$-axis, where positive or negative charges dominate?
In order to gain further understanding in the control of
dissociation asymmetry by the plasmonic field, we simulate the
wave packet dynamics of H$_2^+$ molecule by displacing its
centroid coordinates along the $z$-axis of the nano-structure, as
illustrated in Fig.~4 (top). In a region where positive or
negative charges dominate, the wave packet might be steered
further by attractive or repulsive forces (towards or away from
the metallic surface), respectively.

Figures 4(a)-4(c) show the ionization (Fig.~4(a)) and dissociation
probabilities (Fig.~4(a)) and the asymmetry parameter $A$
(Fig.~4(c)) as a function of centroid offset with respect to the
gap center. Our results show that there is a clear and strong
interplay between ionization, dissociation and the resulting
asymmetry parameter $A$. The asymmetry in ionization probability
is clearly visible in Fig.~4(a). In the negative offset region, a
steep increase in $P_{ion}$ within $-40<$ offset $<0$ is observed
and full ionization occurs beyond this point. In the positive
offset region, on the other hand, $P_{ion}$ increases gradually.
These results suggest that for the negative offset region the
ionized wave packet is pulled towards the metallic surfaces, while
in the positive region is pulled backwards, thus suppressing
ionization as much as possible. After a certain point, however,
the excursion of the wave packet reaches the metallic surface
causing electron absorptions by the metallic
surfaces~\cite{husakou2011theory,yavuz2013gas}.

Wavepacket's dissociation probability $P_{dissoc}$, on the
contrary, reaches a maximum for zero offset and gradually
diminishes in both regions due to the increase of $P_{ion}$. The
dissociation asymmetry parameter $A$ is larger in the gap center
region and maximum, in amplitude, when the offset is $\approx-10$
a.u. Positive values of $A$ in the positive offset region and
negative values of $A$ in the negative offset region is evident
due to an unevenly distribution of charges in the nano-structure
region. In other words, positive offset is causing EL in the upper nuclei, while the negative offset in the
lower one, respectively.

\section{Conclusions and Outlook}

We have studied electron localization (EL) in H$_2^+$ molecules driven
by intense plasmonic fields. These fields are not spatially
homogeneous in the region where the electron wave packet dynamics
takes place and, as a consequence, are thus able to modify
substantially the observables. To illustrate this fact we have solved
the TDSE in reduced dimensions, including both the electron and
nuclear dynamics. This model was proven to be suitable for the
computation of both electron and nuclear related quantities.

We have shown that the spatial inhomogeneous character of the laser
electric field allows us to enhance the localization of the
electron in one of the two heavy ions and given physical grounds
for this behavior. This enhancement can be modified, for instance,
by engineering the geometry of the metal nano-structure.

Furthermore, with our model we can monitor the dissociation
dynamics of the H$_2^+$ molecule as a function of the position
with respect to the center of the gap between the bow-ties. This
analysis is instrumental in order to perform realistic
predictions.

The utilization of plasmonic fields could open the pathway to
perform control of the EL and molecular
dissociation at a more advanced level.

\section{acknowledgment}

I.Y. and Z.A. acknowledge support from BAPKO of MU. Calculations are
performed at the Simulations and Research Lab, Physics Department
of MU. A.C. and M.L. acknowledge the Spanish MINECO project
FOQUS (FIS2013-46768-P), the Catalan AGAUR project SGR 874, EU project QUIC, and ERC AdG OSYRIS.



\end{document}